\documentclass[conference]{IEEEtran}
\usepackage{cite}
\usepackage{amsmath,amssymb,amsfonts}
\usepackage{algorithmic}
\usepackage{graphicx}
\usepackage{textcomp}
\usepackage{xcolor}

\usepackage{tabularx}
\usepackage{longtable} 
\usepackage{colortbl}
\usepackage{color}
\usepackage{tcolorbox} 
\usepackage{stfloats} 
\usepackage{amsthm,thmtools,xcolor} 
\usepackage{orcidlink}

\usepackage{pgfplots}
\pgfplotsset{compat=1.18}
\usetikzlibrary{positioning}
\usepackage{verbatim}
\usepackage{multirow}
\usepackage{multicol}
\usepackage{diagbox}
\usepackage{booktabs}
\usepackage[utf8]{inputenc}
\usepackage{enumitem}
\usetikzlibrary {arrows.meta}
\usetikzlibrary {shapes.arrows}
\usepackage[T1]{fontenc}
\usepackage{tgadventor}
\usetikzlibrary{fit}
\usetikzlibrary{arrows,shapes,decorations,automata,backgrounds,petri}
\usepackage{amsfonts}

\graphicspath{{Images/}} 
\usepackage{eso-pic} 
\usepackage{caption} 
\usepackage{subcaption}
\usepackage{transparent}

\tikzstyle{block} = [draw, rectangle, 
    minimum height=3em, minimum width=6em]
\tikzstyle{sum} = [draw, circle, node distance=1cm]
\tikzstyle{input} = [coordinate]
\tikzstyle{output} = [coordinate]
\tikzstyle{pinstyle} = [pin edge={to-,thin,black}]

\usepackage{fancyhdr}
\fancypagestyle{firstpage}{
  \fancyhf{}

  \fancyhead[L]{\footnotesize This paper has been accepted for publication on IEEE International Conference on Communications (ICC) 2025. The final version published by IEEE is F.F.L. Mariani, M. Zhu and M. Magarini, “Semantic Communications via Features Identification” IEEE ICC 2025 - IEEE International Conference on Communications, Montreal, QC, Canada, 2025.}
  \fancyfoot[L]{\footnotesize\copyright~2025 IEEE. Personal use of this material is permitted. Permission from IEEE must be
obtained for all other uses, in any current or future media, including
reprinting/republishing this material for advertising or promotional purposes, creating new
collective works, for resale or redistribution to servers or lists, or reuse of any copyrighted
component of this work in other works.}
}

\def\BibTeX{{\rm B\kern-.05em{\sc i\kern-.025em b}\kern-.08em
    T\kern-.1667em\lower.7ex\hbox{E}\kern-.125emX}}

\begin{document}

\title{Semantic Communications\\  via Features Identification
}
\author{\IEEEauthorblockN{Federico Francesco Luigi Mariani, Michele Zhu, Maurizio Magarini}
\IEEEauthorblockA{Dipartimento di Elettronica, Informazione e Bioingegneria, Politecnico di Milano, 20133 Milan, Italy \\ 
CORRESPONDING AUTHOR: F. F. L. Mariani, federicofrancesco.mariani@polimi.it}
}
\maketitle
\thispagestyle{firstpage}
\begin{abstract}
The development of the new generation of wireless technologies (6G) has led to an increased interest in semantic communication. Thanks also to recent developments in artificial intelligence and communication technologies, researchers in this field have defined new communication paradigms that go beyond those of syntactic communication to post-Shannon and semantic communication. However, there is still need to define a clear and practical framework for semantic communication, as well as an effective structure of semantic elements that can be used in it. The aim of this work is to bridge the gap between two post-Shannon communication paradigms, and to define a robust and effective semantic communication strategy that focuses on a dedicated semantic element that can be easily derived from any type of message. Our work will take form as an innovative communication method called \textit{identification via semantic features}, which aims at exploiting the ambiguities present in semantic messages, allowing for their identification instead of reproducing them bit by bit. Our approach has been tested through numerical simulations using a combination of machine learning and data analysis. The proposed communication method showed promising results, demonstrating a clear and significant gain over traditional syntactic communication paradigms.
\end{abstract}
%
%
%
\section{Introduction} \label{sec:intro}
\IEEEPARstart{S}{oon} after Shannon’s milestone publication \cite{Amathematicaltheoryofcommunication}, Weaver \cite{weaver_contribution} framed communication as a three-level problem: the technical transmission of symbols, the semantic problem, and the effectiveness problem. The first level, the technical aspect, has been extensively studied and applied in conventional communication systems, which are now approaching the capacity limit defined by Shannon due to the growing demand for a fully connected and integrated world. However, the ultimate scope of communication is often to transfer a specific meaning or intention \cite{strinati20216g}, regardless of the exact wording or symbols used to do so. This goal often clashes with the limitations of physical communication media —such as optical fibers, radio waves, or cables— that are optimized for the transmission of classical, \textit{syntactic}, information.

Semantic communications aim to overcome these limitations by focusing on transmitting only the semantically relevant information for a given task, which can significantly reduce data traffic in future wireless networks. Recently, the development of new communication technologies based on artificial intelligence (AI) has garnered growing interest from both industry and academia \cite{Nokia20246G}, driven by advancements in key enabling technologies such machine learning (ML), large language models (LLM) \cite{bariah2306large}, and statistical modeling. These technologies are expected to play a central role in the upcoming sixth generation (6G) of wireless networks \cite{hoydis2021toward}, where semantic communications and intelligent networks have been identified as core challenges \cite{6g_challenges_and_technologies}.

At the birth of information theory, the semantic aspect of communication was deliberately left out, focusing solely on the mathematical foundation of symbol transmission. Later works, including those by Bar-Hillel and Carnap \cite{carnap1952outline}, sought to introduce preliminary definitions of semantic information within the context of language processing, yet a unified mathematical model that fully captures the nuances of semantic communication has remained elusive. Today, theoretical approaches to semantic communications are many and diverse, such as Ma \textit{et al.} \cite{ma2023theory} defining universal semantic entropy and semantic channel capacity, Zhao \textit{et al.} \cite{zhao2022semantic} \cite{zhao2023joint} modeling the problem using simplicial complex structures, or Kolchinsky \textit{et al.} who studied semantic information via a viability function \cite{kolchinsky2018semantic}. Gaps between these different approaches need to be bridged to harness the full capability of semantic communications.

\textbf{Contributions:} This work aims to consolidate and link the diverse approaches to semantic communications together, and also define a suitable semantic element for next-generation communications that can encapsulate all of the peculiarities of semantic communications and be reasonably practical in its use. Specifically, we investigate a combination of the \textit{Teacher-Apprentice} communication framework \cite{chaccour2022less} with the \textit{message identification} paradigm \cite{cabrera20216g}. Our goal is to propose a new communication method, named "\textit{Identification via semantic features}," which aims to exploit the ambiguity present in semantic communications to identify semantic elements by transmitting their features instead of the whole message. The proposed approach is tested through simulations and numerical analysis to demonstrate its potential for significant savings in transmitted bits, while still achieving high accuracy in identifying semantic elements.

\textbf{Organization:} The paper is structured as follows. Section~\ref{sec:basis} overviews relevant works needed for our research. A detailed description of the proposed communication method is given in Sec.~\ref{sec:Identifiction via semantic features}. A test case with its numerical results is presented in Sec.~\ref{sec:Results} followed by conclusions and further developments in Sec.~\ref{sec:conclusions}.
%
%
%
\section{Semantic architecture and message identification} \label{sec:basis}
The shift in the communication paradigm of semantic communications moves the focus from low-level bit accuracy to higher-level semantic fidelity, which introduces several issues and challenges. Ambiguity is one of the most relevant, especially in natural languages\cite{wasow2005puzzle}, where a message can have different meanings depending on the context of communication. This creates significant challenges for decoding in semantic communication systems, making it difficult, if not impossible. Another major issue is the definition of the structure and atomic element that constitute a semantic message. In classical communication, the structure of a message is predefined in codewords, with the bit serving as its atomic unit. However, in semantic communications, there is no established agreement on the structure or basic unit that constitute a semantic message.

In the following subsections, we will present the key literature that guided our work in addressing these issues, providing the building blocks for our research.
%
%
%
%
\definecolor{torquoise}{RGB}{15,180,180}
\definecolor{lilla}{RGB}{125,157,213}
\definecolor{verde scuro}{RGB}{120,168,120}
\definecolor{blu1}{RGB}{0,155,186}
\definecolor{blu2}{RGB}{0,127,186}
\definecolor{pastelyellow}{rgb}{0.99, 0.99, 0.59}
\definecolor{mossgreen}{rgb}{0.68, 0.87, 0.68}
\definecolor{blue(ncs)}{rgb}{0.0, 0.53, 0.74}
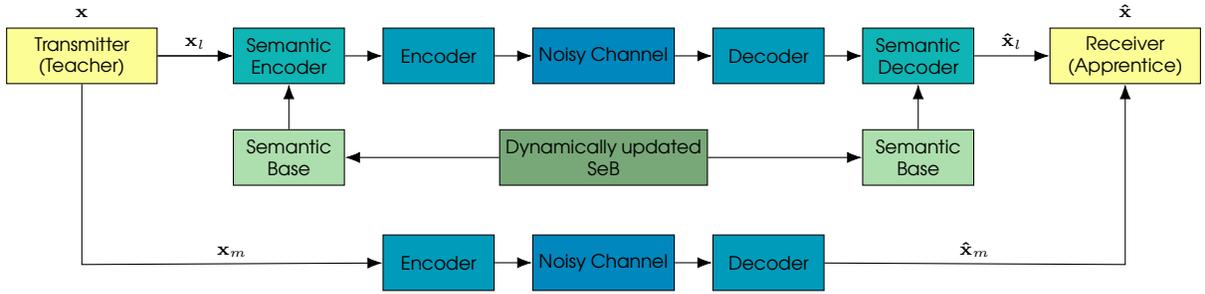
\begin{figure*}[t!]
    \centering
    \Large
    \begin{scriptsize}
        
    {\fontfamily{qag}\selectfont
    \begin{tikzpicture}[node distance = 5mm, scale = 1, transform shape, every text node part/.style={align=center},
                        box/.style = {draw = black, inner sep=5pt,rounded corners=5pt, very thick}
        ]
    
        \node[block, fill = pastelyellow, minimum width = 2cm] (source) {Transmitter\\ (Teacher)};
        \node[above = 0.5mm of source.north](){$\mathbf{x}$};
        \coordinate[below = 2.4cm of source] (source corner);

        \node[block, right = 1cm of source, fill = torquoise](semantic encoder){Semantic\\ Encoder};

        \node[block, right = of semantic encoder, fill = blu1] (encoder) {Encoder};
        
        \node [block, right = of encoder, fill = blue(ncs)] (channel) {Noisy Channel};


        \node[block, right = of channel, fill = blu1] (decoder) {Decoder};

        \node[block, right = of decoder, fill = torquoise] (semantic decoder) {Semantic\\ Decoder};

        \node [block, right = 1cm of semantic decoder, fill = pastelyellow, minimum width = 2cm] (sink) {Receiver\\ (Apprentice)};
        \node[above = 0.5mm of sink.north](){$\mathbf{\hat{x}}$};
        \coordinate[below = 2.4cm of sink](sink corner);

        \node[block, below = 6mm of channel, fill = verde scuro] (knowledge base) {Dynamically updated\\ SeB};

        \node[block, below = 6mm of semantic encoder, fill = mossgreen] (source kb) {Semantic\\  Base};

        \node[block, below = 6mm of semantic decoder, fill = mossgreen] (sink kb){Semantic\\  Base};
        
        \node [block, below =2cm of channel, fill = blue(ncs)] (channel2) {Noisy Channel};

        \node[block, left = of channel2, fill = blu1] (encoder2) {Encoder};


        \node[block, right = of channel2, fill = blu1] (decoder2) {Decoder};
       
        \draw [-{Latex[length=2mm]}] (source) -- node[above] {$\mathbf{x}_l$} (semantic encoder);
        \draw [-{Latex[length=2mm]}] (semantic encoder) -- (encoder);
        \draw [-{Latex[length=2mm]}] (encoder) -- (channel);
        \draw [-{Latex[length=2mm]}] (channel) -- (decoder);
        \draw [-{Latex[length=2mm]}] (decoder) -- (semantic decoder);
        \draw [-{Latex[length=2mm]}] (source) -- (semantic encoder);
        \draw [-{Latex[length=2mm]}] (semantic decoder) -- node[above] {$\mathbf{\hat{x}}_l$} (sink);


        \draw [-{Latex[length=2mm]}] (knowledge base) -- (source kb);
        \draw [-{Latex[length=2mm]}] (knowledge base) -- (sink kb);
        \draw [-{Latex[length=2mm]}] (source kb) -- (semantic encoder);
        \draw [-{Latex[length=2mm]}] (sink kb) -- (semantic decoder);

        \draw [-{Latex[length=2mm]}] (encoder2) -- (channel2);
        \draw [-{Latex[length=2mm]}] (channel2) -- (decoder2);

        \draw [-{Latex[length=2mm]}] (source) -- (source corner) -- node[above] {$\mathbf{x}_m$} (encoder2);
        \draw [-{Latex[length=2mm]}] (decoder2) -- node[above] {$\mathbf{\hat{x}}_m$} (sink corner) -- (sink);
        
    \end{tikzpicture}
    }
    \caption{Block diagram of the hybrid semantic communication network scheme with separated learnable (top) and memorizable (bottom) data streams. The learnable data path is dedicated to the transmission of structured data $\mathbf{x}_l$ which will carry semantic information, while the memorizable data path is dedicated to the transmission of unstructured data $\mathbf{x}_m$ that is going to complement the semantic one.\vspace{-4mm}}
    \label{fig:semantic hybrid scheme}
    \end{scriptsize}
\end{figure*}
\subsection{Semantic Communication Architecture} \label{sec:semantic architecture}
The definition of a semantic communication framework helps us addressing the problems of ambiguity and the need to define a semantic element's structure. Focusing on the physical layer, as illustrated by Zhang \textit{et al.} \cite{zhang2022toward}, a semantic communication framework includes several critical components that differentiate it from classical communication architectures. These include: the \textbf{semantic base} (SeB), which is responsible for establishing a common communication background between the transmitter and receiver, addressing part of the ambiguity problem; and the \textbf{semantic encoder/decoder pair}, responsible for encoding and decoding raw data inputs into semantic information. In semantic communications, the transmitter and receiver each possess a local knowledge base, which may differ at any given time. To address this, a third element — the \textbf{dynamically updated SeB} — must be introduced to synchronize the knowledge bases, allowing coherent semantic decoding at the receiver's side.

Building on these elements, Chaccour \textit{et al.} \cite{chaccour2022less} redefined the communication architecture by introducing the concept of a \textit{hybrid semantic-syntactic network}. This architecture splits the information carried by a message into two components: a \textbf{learnable data stream} $\mathbf{x}_l$, which represents patterns and structures that can be exploited to make transmission more efficient, and a \textbf{memorizable data stream} $\mathbf{x}_m$, which depends on random components that cannot be represented by means of patterns or data structures and is more suitable for classical syntactic communication.

In this framework, two agents are defined: The \textbf{teacher} which replaces the transmitter, and whose role is to split the data stream $\mathbf{x}$ into $\mathbf{x}_l$ and $\mathbf{x}_m$ while synchronizing the SeB with the receiver; and the \textbf{apprentice}, which takes the role of the receiver, who learns data structures from the teacher and uses them either to reconstruct the original message or to generate new messages. This hybrid semantic communication scheme, based on a teacher-apprentice architecture, is illustrated in Fig.~\ref{fig:semantic hybrid scheme}. From a practical point of view, there is no reason for the actual physical communication to happen in any way other than a syntactic one. Therefore a syntactic communication channel is still present also in the semantic communication line. Errors introduced by it can be mitigated through signal processing techniques, such as matched filtering, equalization, the Nyquist criterion, and pulse shaping; error control techniques, such as forward error correction and interleaving; modulation and coding techniques; and diversity techniques, including time, frequency, and space diversity~\cite{barry2012digital}. 
%
%
%
\definecolor{amber(sae/ece)}{rgb}{1.0, 0.49, 0.0}
\definecolor{amethyst}{rgb}{0.6, 0.4, 0.8}
\definecolor{non-photoblue}{rgb}{0.64, 0.87, 0.93}
\definecolor{napiergreen}{rgb}{0.16, 0.5, 0.0}
\definecolor{mossgreen}{rgb}{0.68, 0.87, 0.68}
\definecolor{blue(ncs)}{rgb}{0.0, 0.53, 0.74}
\definecolor{amber}{rgb}{1.0, 0.75, 0.0}
\definecolor{airforceblue}{rgb}{0.36, 0.54, 0.66}
\begin{figure*}[t!]
    %
    %
    %
    %
    \hspace{-5cm}
    \begin{subfigure}[t]{\textwidth}
    \centering
    \Large
    \includegraphics[]{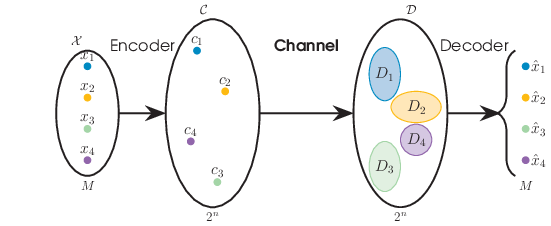}
    \caption{}
    \label{fig:Shannon com paradigm}
    \end{subfigure}
    \hspace{-9.3cm}
    %
    %
    %
    %
    \begin{subfigure}[t]{\textwidth}
        \centering
        \Large
        \includegraphics[]{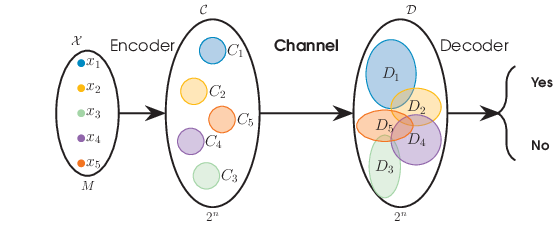}
        \caption{}
    \label{fig:Identification com paradigm}
    \end{subfigure}
    \caption{Shannon's communication paradigm in (a) compared to identification paradigm in (b), as illustrated in \cite{cabrera20216g}. In (a) the receiver needs to the decode the received distorted code-word $c_i$ and retrieve an estimate of the transmitted message $\hat{x}_i$. In (b) the receiver needs to determine if the received distorted codewords set $C_i$ encodes a message that is relevant to him or not.\vspace{-4mm}}
    \label{fig:Shannon vs Identification}
\end{figure*}
%
%
%
%
\begin{figure}[!b]
    \centering
    \includegraphics[]{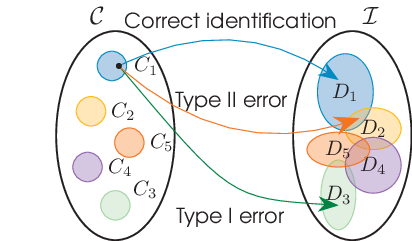}
    \caption{Error types of message identification as in \cite{cabrera20216g}. Type I errors represent the syntactic communication errors, leading to strong incorrect identification. Type II errors are created by the ambiguity present in the message identification paradigm and can lead to incorrect identification.}
    \label{fig:error type}
\end{figure}
\subsection{Message Identification} \label{sec:message identification}
Given the different requirements of semantic communication, the communication paradigm is shifted in \cite{cabrera20216g} from a syntactic approach to a post-Shannon identification paradigm. The theory of identification via channels was first introduced by Ahlswede and Dueck in \cite{ahlswede1989identification}. Here, the receiver’s primary task is to determine whether the transmitted message is relevant to him or not, without needing to reconstruct it bit-by-bit. Furthermore, no prior information is available regarding which messages are relevant to the receiver. To achieve this, the transmitter encodes messages in such a way that the receiver can reliably identify those of interest. Randomized identification (RI) codes were used in this context with significant success. It has been shown in \cite{ahlswede1989identification} that RI codes allow the identification of up to $2^{2^{nC}}$ messages, in contrast to the $2^{nC}$ messages that can be decoded in a traditional Shannon-like paradigm \cite{Amathematicaltheoryofcommunication} \cite{wolfowitz2012coding}, where $n$ is the block length and $C$ is the channel capacity defined by Shannon. 

In Fig.~\ref{fig:Shannon com paradigm}, Shannon’s communication paradigm is illustrated. In this model, a message $x_i$, taken from a set $\mathcal{X}$ of $M$ possible messages, is encoded into a single codeword $c_i \in \mathcal{C}$, which is then transmitted over a noisy channel. At the receiver side, the codeword, which has been distorted by the channel, creates a decoding region $D_i \subset \mathcal{D}$. The task of the receiver is to retrieve an estimate $\hat{x}_i$ of the original message from $D_i$. Figure~\ref{fig:Identification com paradigm} depicts the message identification paradigm, where the aim is to determine if the received message is relevant to the receiver or not. The transmitter encodes the message $x_i$ into a codeword belonging to a set of possible codewords $C_i \subset \mathcal{C}$, where the number of elements present in the set $C_i$ depends on the coding technique used. These sets differ by at least $d$ elements from each other, where $d$ is known as the Hamming distance. Codewords are then transmitted over a noisy channel which distorts them, creating the decoding regions $D_i$ at the receiver's side. These regions may now overlap with each other, resulting in three possible outcomes after transmission: \textbf{Correct identification}, \textbf{missed identification (Type I error)}, or \textbf{false identification (Type II error)} as illustrated in Fig.~\ref{fig:error type}. 
\textbf{Type I} errors, represent classical communication errors caused by noise introduced by the channel and can be mitigated using standard communication techniques. Therefore, they will not be considered in this work, where we assume to be able to deal with them in both the memorizable and learnable data paths presented in Fig.\ref{fig:semantic hybrid scheme}.
\textbf{Type II} errors, on the other hand, are particularly relevant in the context of semantic communications, as they introduce ambiguity, which impairs the identification of relevant messages. Our scope will be that of exploiting this ambiguity to identify the transmitted messages in an efficient manner.

Shifting the communication paradigm to message identification offers the advantage of identifying a larger set of messages with the same number of transmitted bits. Using the message identification paradigm with RI codes \cite{ahlswede1989identification}, up to $2^{2^{nC}}$ messages can be identified by exploiting ambiguity carried by \textbf{Type II} errors, representing a considerable improvement over Shannon’s original approach. Additionally, applying this coding strategy helps us with the definition of the structure of semantic elements and their atomic units later in this work.
%
\section{Identification via Semantic Features.} \label{sec:Identifiction via semantic features}
In this section, we introduce our contribution: an innovative semantic communication method called \textit{Identification via Semantic Features}. Based on the network architecture described in \cite{chaccour2022less} and applying the identification principles outlined in \cite{cabrera20216g}, our goal is to exploit ambiguity in order to identify the same number of messages of a Shannon communication paradigm with a significant reduction in the number of transmitted bits. 
\subsection{Semantic Representation}
Consider now a message set $\mathcal{X}$ with cardinality: $|\mathcal{X}| = K$. Each message in the set is a vector that can be expressed as
\begin{equation}
    \mathbf{x} = \{\mathbf{x}_{m}, \mathbf{x}_{l}\} \in \mathcal{X},
\end{equation}
where $\mathbf{x}_{m}$ and $\mathbf{x}_{l}$ are the memorizable and learnable parts that compose the message $\mathbf{x}$, respectively. We now wish to associate its learnable part $\mathbf{x}_{l}$ to an identity vector $\mathbf{i} \in \mathcal{I}$, where $\mathcal{I}$ is the set of all possible identities. The relation between learnable data and its identity can be expressed as \begin{equation}
    \mathbf{i} = f_E(\mathbf{x}_{l}) \in \mathcal{I},
\end{equation}
where $f_E(\mathbf{x}_{l})$ is a function that takes the learnable data points $\mathbf{x}_{l}$ as input and associates them to a corresponding identity $\mathbf{i}$. 
Each identity is a vector composed by $n$ feature elements $f(n)$, each with size $q$ bits. A set of features defines an identity as
\begin{equation}
    \mathbf{i} = \{ f(1), f(2), \dots, f(n) \} \in \mathcal{I},
\end{equation}
as illustrated in Fig.~\ref{fig:feature mapp}. The set $\mathcal{I}$ contains at most as many identities as there are data points in $\mathcal{X}$, so that 
\begin{equation}
    |\mathcal{I}| \leq |\mathcal{X}| = K.
\end{equation}

\definecolor{pastelred}{rgb}{1.0, 0.41, 0.38}
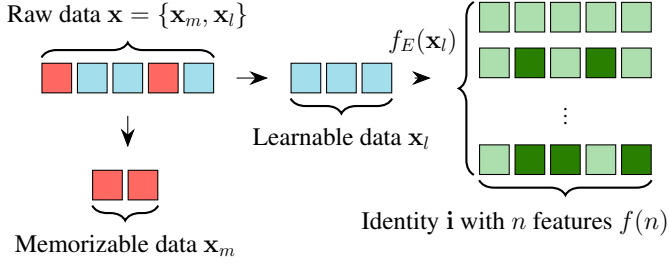
\begin{figure}[!t]
    \centering
    \Large
    \begin{tikzpicture}[
        squarenode/.style={rectangle, draw=black, minimum size=.58cm,font=\scriptsize,>=latex'},
        scale = 0.68, transform shape
    ]
        \node[squarenode, fill = mossgreen] (c11) {};
        \node[squarenode, fill = mossgreen] (c12) [right = 1mm of c11] {};
        \node[squarenode, fill = mossgreen] (c13) [right = 1mm of c12] {};
        \node[squarenode, fill = mossgreen] (c14) [right = 1mm of c13] {};
        \node[squarenode, fill = mossgreen] (c1n) [right = 1mm of c14] {};

        \node[squarenode, fill = mossgreen] (c21) [below = 3mm of c11] {};
        \node[squarenode, fill = napiergreen] (c22) [right = 1mm of c21] {};
        \node[squarenode, fill = mossgreen] (c23) [right = 1mm of c22] {};
        \node[squarenode, fill = napiergreen] (c24) [right = 1mm of c23] {};
        \node[squarenode, fill = mossgreen] (c25) [right = 1mm of c24] {};
        
        \node[squarenode, fill = mossgreen] (c31) [below = 1.3cm of c21] {};
        \node[squarenode, fill = napiergreen] (c32) [right = 1mm of c31] {};
        \node[squarenode, fill = napiergreen] (c33) [right = 1mm of c32] {};
        \node[squarenode, fill = mossgreen] (c34) [right = 1mm of c33] {};
        \node[squarenode, fill = napiergreen] (c35) [right = 1mm of c34] {};

        \node[squarenode, fill = non-photoblue] (l1) [left = 2cm of c21.south] {};
        \node[squarenode, fill = non-photoblue] (l2) [left = 1mm of l1] {};
        \node[squarenode, fill = non-photoblue] (l3) [left = 1mm of l2] {};
        
        \node[squarenode, fill = non-photoblue] (m1) [left = 1.5cm of l3] {};
        \node[squarenode, fill = pastelred] (m2) [left = 1mm of m1] {};
        \node[squarenode, fill = non-photoblue] (m3) [left = 1mm of m2] {};
        \node[squarenode, fill = non-photoblue] (m4) [left = 1mm of m3] {};
        \node[squarenode, fill = pastelred] (m5) [left = 1mm of m4] {};

        \node[squarenode, fill = pastelred] (xm1) [below = 1.5cm of m3.south east] {};
        \node[squarenode, fill = pastelred] (xm2) [left = 1mm of xm1] {};
        
        \node (input) [right = 1mm of m1] {};
        \node (input2) [right = 1mm of l1] {};
        \node (output2) [left = 9mm of c21.south] {};
        \node (output) [left = 1mm of l3] {};
        \node (x_min) [below = 1mm of m3] {};
        \node (x_mout) [above = 1mm of xm1.north west] {};
        \node (dots) [below = 2mm of c23] {\large $\vdots$};
        
        \draw[-{Stealth[length=2.5mm]}] (input) -- (output);
        \draw[-{Stealth[length=2.5mm]}] (input2) -- node[above = 3mm] {$f_E(\mathbf{x}_l)$} (output2);
        \draw[-{Stealth[length=2.5mm]}] (x_min) -- (x_mout);
        
        \draw[thick,decorate,decoration={brace,raise=2pt,amplitude=2mm}] (m5.north west) -- node[above = 5mm] {Raw data $\mathbf{x} = \{\mathbf{x}_m, \mathbf{x}_l\}$} (m1.north east);
        \draw[thick,decorate,decoration={brace,raise=2pt,mirror,amplitude=2mm}] (c31.south west) -- node[below left = 5mm and -22mm] {Identity $\mathbf{i}$ with $n$ features $f(n)$} (c35.south east);
        \draw[thick,decorate,decoration={brace,raise=2pt,mirror,amplitude=2mm}] (c11.north west) -- (c31.south west);
        \draw[thick,decorate,decoration={brace,raise=2pt,mirror,amplitude=2mm}] (l3.south west) -- node[below = 5mm] {Learnable data $\mathbf{x}_l$} (l1.south east);
        \draw[thick,decorate,decoration={brace,raise=2pt,mirror,amplitude=2mm}] (xm2.south west) -- node[below = 5mm] {Memorizable data $\mathbf{x}_m$} (xm1.south east);
    \end{tikzpicture}
    \caption{Learnable data gets separated from memorizable data and mapped into an identity vector. Different shades of green represent different feature's value between identities.}
    \label{fig:feature mapp}
    \vspace{-.15cm}
\end{figure}

\begin{figure}[!t]
    \centering
    \includegraphics[]{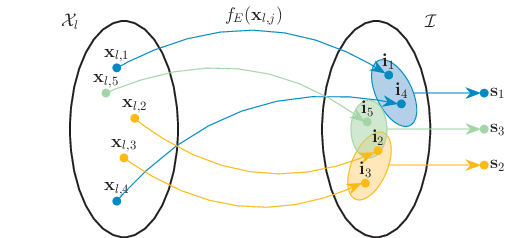}
    \caption{Mapping learnable data points $\mathbf{x}_{l,i}$ to their corresponding identities $\mathbf{i}_i$ and then in to semantic elements $\mathbf{s}_k$.}
    \label{fig:mapp}
    \vspace{-.18cm}
\end{figure}

Thus, each identity corresponds to at least one learnable data point:
\begin{equation}
    \forall \mathbf{x}_{l} \in \mathcal{X}, \exists \mathbf{i} \in \mathcal{I} \mid f_E(\mathbf{x}_{l}) = \mathbf{i}.
\end{equation}
Making $f_E(\mathbf{x}_{l})$ a subjective function, as illustrated in Fig.~\ref{fig:mapp}.

Features in the same position represent the same aspect of the data, meaning it is possible for two different identities $\mathbf{i}_{i} \neq \mathbf{i}_{j}$, to share the same feature value: $f_i(k) = f_j(k)$. However, this does not create an issue since the ordered sequence of features uniquely describes the message, giving rise to the concept of \textit{identity}. If instead two messages share all their features, they will be associated with the same identity:
\begin{equation}
    f_E(\mathbf{x}_{l,i}) = f_E(\mathbf{x}_{l,j}) \Rightarrow \mathbf{i}(\mathbf{x}_{l,i}) = \mathbf{i}(\mathbf{x}_{l,j}) = \mathbf{i}.
\end{equation}
This behavior fits well within the concept of ambiguity while retaining the ability to distinguish messages at the receiver’s end. Additionally, messages with the same semantic content will typically share similar feature values. This allows us to recognize messages with matching semantic content $\mathbf{s}$ because their identities $\mathbf{i}$ will be close to each other in the identity space $\mathcal{I}$. The semantic element can thus be attributed to a subset of similar identities $\mathcal{I}^{\prime}$, represented as: 
\begin{equation}
    \mathbf{s} = \mathcal{I}^{\prime} \subset \mathcal{I},
\end{equation}
as shown in Fig.~\ref{fig:mapp}. The semantic element will therefore represent the structure of features that is common to all identities in a given subset $\mathcal{I}^{\prime}$, giving a blueprint for all the identities belonging to $\mathcal{I}^{\prime}$.  
\definecolor{blue(ncs)}{rgb}{0.0, 0.53, 0.74}
\definecolor{amber}{rgb}{1.0, 0.75, 0.0}
\definecolor{airforceblue}{rgb}{0.36, 0.54, 0.66}

\subsection{Features Selection and Forward}
As discussed in Sec.~\ref{sec:semantic architecture}, once all messages have been associated with their identities, the teacher can identify subsets of identities that correspond to specific semantic elements and transmit this knowledge to the apprentice, providing him with an updated SeB. Following \cite{cabrera20216g}, in our approach, the teacher randomly selects a feature to transmit from the identity associated with the data point he wants to transmit. We assumed random selection to be the best option if no prior information is available at the transmitter or receiver side. The selected feature is then encapsulated in a packet, containing both the feature’s value and its position. This packet is then transmitted to the apprentice. We assume the apprentice to have no prior information about which message the teacher is sending, therefore it will use the received packet to identify which message the transmitter is trying to communicate.

Upon receiving the packet, the apprentice compares the received feature with those in the same position belonging to identities present in his SeB. Since each identity is linked to a semantic element, the apprentice first identifies the identity and then the corresponding semantic element. To complete the identification process, more than one packet is needed. This means that before selecting a semantic element, the process of sending packets continues until the probability of having a specific semantic element given the received chain of features exceeds a certain confidence threshold $\lambda$ to be set. At this point, the apprentice interrupts the transmission and selects as the semantic element the most probable one, retrieving the estimated identity from his SeB. This aspect will be discussed later on in the Sec.~\ref{sec:3.3}.

\begin{figure}[h!]
    \centering
    \Large
    \begin{tikzpicture}[
        squarenode/.style={rectangle, draw=black, minimum size=.5cm,font=\scriptsize,>=latex'},
        scale = 0.65, transform shape
    ]

        \node[squarenode, fill = mossgreen] (i11) {};
        \node[squarenode, fill = mossgreen] (i12) [right = 1mm of i11] {};
        \node[squarenode, fill = mossgreen] (i13) [right = 1mm of i12] {};
        \node[squarenode, fill = pastelred] (i14) [right = 1mm of i13] {};
        \node[squarenode, fill = mossgreen] (i1n) [right = 1mm of i14] {};

        \node[squarenode, fill = mossgreen] (i21) [below = 3mm of i11] {};
        \node[squarenode, fill = napiergreen] (i22) [right = 1mm of i21] {};
        \node[squarenode, fill = mossgreen] (i23) [right = 1mm of i22] {};
        \node[squarenode, fill = napiergreen] (i24) [right = 1mm of i23] {};
        \node[squarenode, fill = mossgreen] (i2n) [right = 1mm of i24] {};
        
        \node[squarenode, fill = mossgreen] (i31) [below = .8cm of i21] {};
        \node[squarenode, fill = mossgreen] (i32) [right = 1mm of i31] {};
        \node[squarenode, fill = napiergreen] (i33) [right = 1mm of i32] {};
        \node[squarenode, fill = napiergreen] (i34) [right = 1mm of i33] {};
        \node[squarenode, fill = napiergreen] (i3n) [right = 1mm of i34] {};

        \node[fill = pastelred, single arrow, rotate = -90] (arrow) [above left = .65cm and .45cm of i14.north] {\footnotesize \fontfamily{qag}\selectfont \textbf{Match}};

        \node[draw = black, minimum size = 1.3cm, font = \large, left = 2cm of i21.south] (packet) {\textbf{(4, $f(4)$)}};

        \node[squarenode, fill = pastelred] (xm1) [below  = 1.3cm of packet] {};
        \node[squarenode, fill = pastelred] (xm2) [right  = 1mm of xm1] {};
        
        \node[squarenode, fill = non-photoblue] (s1) [right = 1.5cm of i1n] {};
        \node[squarenode, fill = non-photoblue] (s2) [right = 1mm of s1] {};
        \node[squarenode, fill = non-photoblue] (s3) [right = 1mm of s2] {};

        \node (packet1) [right = .2mm of packet] {};
        \node (identity) [left = 4mm of i21.south] {};

        \node (identity1) [right = .2mm of i1n] {};
        \node (semantic) [left = .2mm of s1] {};

        \node (mem) [right = .2mm of xm2] {};
        \node (memsum) [below = 5.05cm of s2] {};

        \node (sum) [circle, draw = black,  below = .8cm of s2] {\Large $+$};
        \coordinate[below = 1.25cm of sum] (corner);
        
        \node (dots) [below = 0mm of i23] {\large $\vdots$};

        \node[squarenode, fill = pastelred] (x1) [right = 1cm of sum] {};
        \node[squarenode, fill = non-photoblue] (x2) [right = 1mm of x1] {};
        \node[squarenode, fill = non-photoblue] (x3) [right = 1mm of x2] {};
        \node[squarenode, fill = pastelred] (x4) [right = 1mm of x3] {};
        \node[squarenode, fill = non-photoblue] (x5) [right = 1mm of x4] {};

        \node (x^) [left = .1mm of x1] {};
        
        \draw[-{Stealth[length=2.5mm]}] (packet1) -- node[above = 2cm] {\fontfamily{qag}\selectfont Comparison} (identity);
        \draw[-{Stealth[length=2.5mm]}] (identity1) -- node[above = 1cm] {\fontfamily{qag}\selectfont Selection}(semantic);
        \draw[-{Stealth[length=2.5mm]}] (mem) -- (corner) -- (sum.south);
        \draw[-{Stealth[length=2.5mm]}] (s2.south) -- (sum);
        \draw[-{Stealth[length=2.5mm]}] (sum) -- (x^);
        
        \draw[thick,decorate,decoration={brace,raise=2pt,mirror,amplitude=2mm}] (i11.north west) -- (i31.south west);
        \draw[thick,decorate,decoration={brace,raise=2pt, mirror, amplitude=2mm}] (s3.north east) -- node[above = 5mm] {$\hat{\mathbf{s}}$} (s1.north west);
        \draw[thick,decorate,decoration={brace,raise=2pt, mirror, amplitude=2mm}] (xm1.south west) -- node[below = 5mm] {$\hat{\mathbf{x}}_m$} (xm2.south east);
        \draw[thick,decorate,decoration={brace,raise=2pt, mirror, amplitude=2mm}] (x1.south west) -- node[below = 5mm] {$\hat{\mathbf{x}} = \{ \hat{\mathbf{x}}_m ; \hat{\mathbf{s}}$\}} (x5.south east);
        
    \end{tikzpicture}
    \caption{Upon receiving a packet, the receiver compares the received feature's value with the ones present in its knowledge base and selects an estimated semantic element $\hat{\mathbf{s}}$. It then composes it with the received memorizable data $\hat{\mathbf{x}}_m$ and recovers an estimate of the transmitted message $\hat{\mathbf{x}}$.\vspace{-4mm}}
    \label{fig:message reconstruction}
\end{figure}
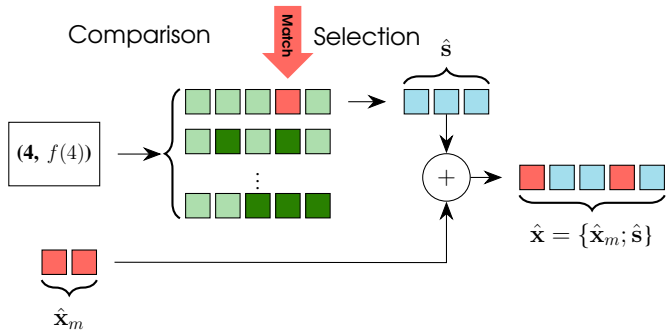

Once the apprentice has identified the semantic element, it can combine it with the memorizable data $\mathbf{x}_m$, which has been transmitted syntactically, to reconstruct an estimate $\hat{\mathbf{x}}$ of the original message:
\begin{equation}
    \hat{\mathbf{x}} = \{ \hat{\mathbf{x}}_{m}, \hat{\mathbf{s}} \},
\end{equation}
where $\{ \hat{\mathbf{x}}_{m}, \mathbf{\hat{s}} \}$ represents the composition of the estimated memorizable data $\hat{\mathbf{x}}_\mathbf{m}$ and the estimated semantic element $\mathbf{\hat{s}}$, as illustrated in Fig.~\ref{fig:message reconstruction}.
\subsection{Semantic Element Identification} \label{sec:3.3}
Identities corresponding to the same semantic object are close to each other in the $n$-dimensional space $\mathcal{I}$. Consequently, we represent each semantic element $\mathbf{s}$ with a reference identity $\bar{\mathbf{i}}$, calculated as the mean identity vector of the identities belonging to a given semantic element: 
\begin{equation}
    \bar{\mathbf{i}} = \mathbb{E}\{\mathbf{i}\}, \forall \mathbf{i} \in \mathbf{s}.
\end{equation}
To identify the correct semantic element based on the received packets, we need to compute the probability: 
\begin{equation} \label{eq:}
    P(\mathbf{s} | \mathbf{f}_{t}),
\end{equation}
where $\mathbf{f}_{t}$ represents the feature vector composed of the received feature values up to time $t$. We note that the correspondence between $\bar{\mathbf{i}}$ and $\mathbf{s}$ is one-to-one:
\begin{equation}\label{eq:i_k=s_k}
\hspace{-2mm}
     P(\bar{\mathbf{i}} | \mathbf{f}_{t}) = P(\mathbf{s} | \mathbf{f}_{t}).
\end{equation}
Finding the argument that maximises this probability let us retrieve the most probable $\bar{\mathbf{i}}$, and consequently the most probable $\hat{\mathbf{s}}$: 
\begin{equation}\label{eq:S_k}
    \hat{\mathbf{s}} = \arg \max_{\bar{\mathbf{i}}} \{ P(\bar{\mathbf{i}} | \mathbf{f}_{t}) \}.
\end{equation}
For more than one feature, computing this probability becomes increasingly complex as the number of features grows. Joint probabilities between different features must be taken into account, and these are generally not independent, making the task computationally demanding. To approximate this probability, we use the \textit{Inverse Distance Weighting} (IDW) algorithm \cite{franke1980smooth}, which attributes a weight coefficient to a point in space based on its distance from a set of reference points. The set of weights is calculated as: 
\begin{equation}\label{eq: w_i}
    \mathbf{w} = \frac{ \left( \frac{R-\mathbf{d}}{R \mathbf{d}} \right) ^2}{\sum_i{ \left( \frac{R-d_i}{Rd_i} \right) ^2}},
\end{equation}
where $R$ is the maximum distance in the set of points and $d_i$ is the Euclidean distance $d_i = \sqrt{(x-x_i)^2 + (y-y_i)^2+ \dots}$ and $\mathbf{d}$ is the vector of euclidean distances from all the reference identities. By applying IDW, we are considering all the features at the same time, thus taking \textit{context} into account. As more features are received, the probability $P(\mathbf{\mathbf{s}}_t|\mathbf{f}_t)$ needs to be updated accordingly as
\begin{equation}\label{eq:P(S|(F...))}
    P(\mathbf{\mathbf{s}}_t|\mathbf{f}_t) = 
    \frac{P(\mathbf{\mathbf{s}}_{t-1}|\mathbf{f}_{t-1}) + \mathbf{w}_{t}}{|P(\mathbf{\mathbf{s}}_{t-1}|\mathbf{f}_{t-1}) + \mathbf{w}_{t}|},  
\end{equation}
where $\mathbf{f}_t = [f_t, f_{t-1}, \dots]$ represents the collected features up to time $t$, which are drawn from a random permutation of features of $\mathbf{i}$, and $\mathbf{w}_{t}$ represents the probability weights vector calculated as in (\ref{eq: w_i}) at time $t$. Note that the dimensionality of $d_i$ increases with the number of received features, and only the features of $\mathbf{i}$ that match the received ones are considered. 
Once the probability in (\ref{eq:P(S|(F...))}) reaches the confidence threshold $\lambda$, i.e. $ P(\mathbf{\mathbf{s}}_t|\mathbf{f}_t) \geq \lambda$, the transmission can stop and the most probable semantic element can be selected according to (\ref{eq:S_k}). The value $\lambda$ represents the lower probability bound for which a semantic element is eligible to be considered as the correct one from the receiver.
%
%
%
\vspace{-1mm}
\section{Numerical Results} \label{sec:Results}
In this section, the results of the simulation used to test our proposed method are presented. Mainly due to the rich background in AI and ML classification techniques, we chose an image classification scenario. In our scenario the receiver needs to identify the class of each image sent using the principle highlighted in Sec.~\ref{sec:Identifiction via semantic features}. The transmitter will send packets of features taken from the feature vector generated by a convolutional neural network (CNN). The feature vector is generated as the output before the last dense layer of the network. The features vector generated in this way is considered to be the learnable part associated with each image. Separation between learnable and memorizable parts is not considered in this simulation.

Our goal is to identify the learnable part, i.e. the features vector, with the lowest possible number of transmitted bits. The chosen image dataset is Imagenette \cite{Howard_Imagenette_2019}, derived from the well-known ImageNet dataset \cite{Imagenet} as a subset of ten classes. For classification and for computing the features vector, we used the ResNet50 model \cite{he2016deep}, pre-trained on the ImageNet dataset and then fine-tuned through transfer learning \cite{zhuang2020comprehensive} and specialized for the Imagenette dataset. The simulation was performed testing different feature dimensions in the range  $[2^5,2^{11}]$ to construct the identities, including a specific case with only five features to examine scenarios where the number of features is smaller than the number of classes. We varied $\lambda$ between $10\%$ and $100\%$, with $10\%$ representing the worst-case scenario. The transmission stops once the probability  $P(\mathbf{\mathbf{s}}_t|\mathbf{f}_t)$ exceeds $\lambda$, and the receiver selects the most probable semantic element as in (\ref{eq:S_k}).

\begin{figure}[!t]
    \centering
    \includegraphics[width = .98\linewidth]{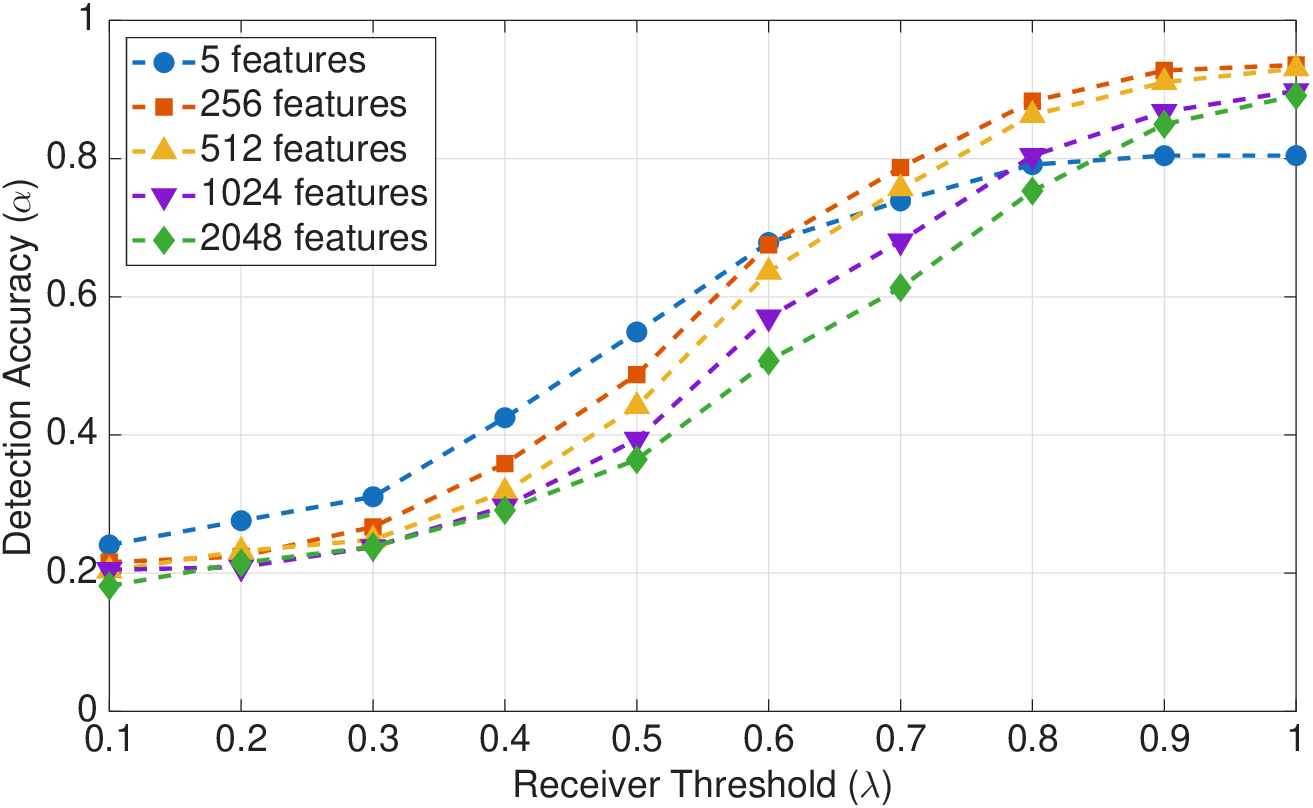}
    \caption{Identification accuracy $\alpha$ with respect to receiver threshold $\lambda$ for different number of features used. \vspace{-3mm}}
    \label{fig:accuracy}
\end{figure}

The following performance indicators were used:

\textbf{Identification Accuracy}, which is defined as
\begin{equation}
    \alpha(\lambda) = 1 - \frac{N_\varepsilon(\lambda)}{N_i},
\end{equation}
where $N_{\varepsilon}(\lambda)$ represents the number of identification errors during transmission for a given $\lambda$ and $N_i$ is the total number of images in the dataset. This metric reflects the accuracy with which the receiver identifies the correct semantic element. Figure~\ref{fig:accuracy} shows the results gathered from our simulations. We observe that, starting from the worst case scenario where identities are equiprobable, identification accuracy decreases as the number of features used to create the identity vectors increases, highlighting the \textit{"curse of dimensionality"} \cite{Zaki_Meira_jr}. In fact, with more features, we converge to lower values of identification accuracy. This is because by increasing the number of dimensions in the space $\mathcal{I}$, the distances $d_i$ evaluated in (\ref{eq: w_i}) become more uniformly distributed, therefore less selective for identification purposes. 

\begin{figure}[!t]
    \centering
    \includegraphics[width= .98\linewidth]{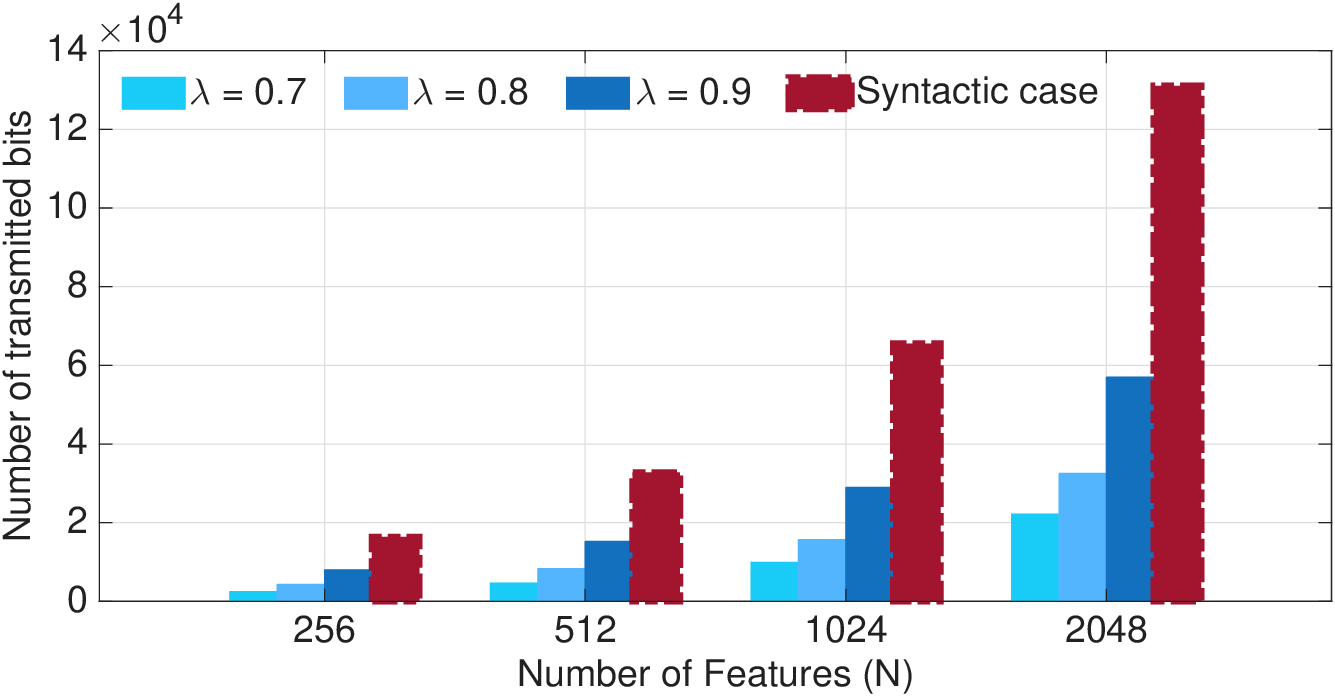}
    \caption{Number of transmitted bits, comparison between semantic and syntactic approach at the vary of receiver threshold.\vspace{-3mm}}
    \label{fig:BER2}
\end{figure}

\begin{figure}[!t]
    \centering
    \includegraphics[width = .98\linewidth]{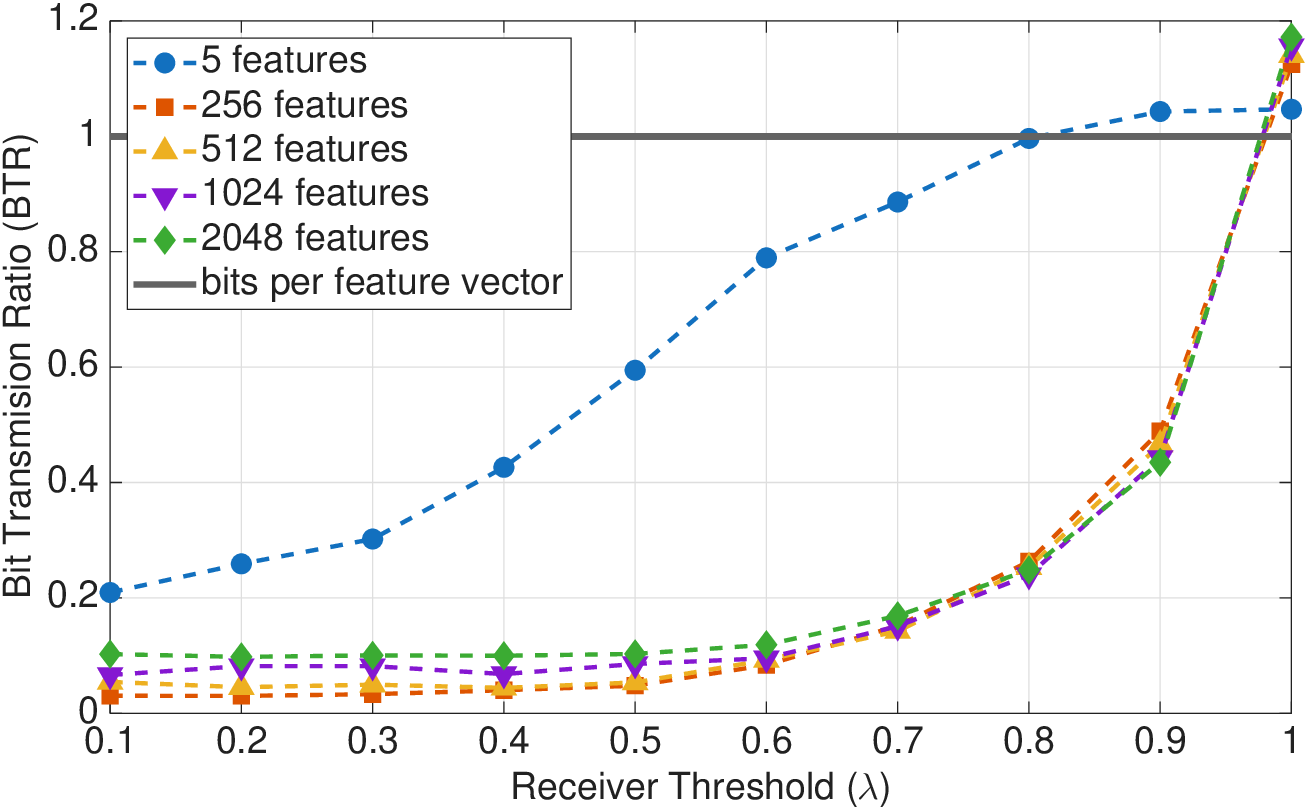}
    \caption{Bit transmission ratio for different number of features used. The black horizontal line represents the limit where we are transmitting the same amount of bits as the syntactic case.\vspace{-3mm}}
    \label{fig:BTR}
\end{figure}

\textbf{Bit Transmission Ratio (BTR)}, which measures the ratio of bits transmitted using our semantic approach compared to a traditional syntactic method for the same identity vector $\mathbf{i}$. The BTR is calculated as:
\begin{equation}\label{eq:BTR}
    \mathrm{BTR}(\lambda) = \frac{(\lceil\log_2(N)\rceil + q) \cdot N_{p}(\lambda)}{q \cdot N},
    \vspace{-1mm}
\end{equation}
where $N_{p}(\lambda)$ is the number of packets received before the end of transmission, $q$ is the number of bits used to represent the feature's value, and $N$ is the total number of features. Given that the position of each feature is encoded with $\lceil\log_2(N)\rceil$ bits, we have a total of $\lceil\log_2(N)\rceil $$\,+\,$$ q$ bits per packet, which leads to $(\lceil\log_2(n)\rceil +q) \cdot N$ bits to transmit all the $N$ features packetized as described in Sec.~\ref{sec:Identifiction via semantic features}. Instead, only $q$$\,\cdot\,$$N$ bits are needed to simply transmit the entire identity vector in a syntactic fashion. In what follows, we set $q$$\,=\,$$64$. Figure~\ref{fig:BER2} demonstrates the bit-saving advantage of the semantic approach. It compares the number of transmitted bits using our semantic approach to the number of bits that would have been required for the syntactic transmission of the entire features vector. For example, $\mathrm{BTR}$$\,=\,$$0.5$ means that we are transmitting $50\%$ of the bits required in the syntactic case. Interestingly, Fig.~\ref{fig:BTR} shows that having more features results in a better BTR for higher values of $\lambda$, despite a marginal decrease in identification accuracy. This is because BTR is a ratio, where the denominator of (\ref{eq:BTR}) increases linearly with $N$, while the numerator is driven by $\log_2(N)$. For the considered values, (\ref{eq:BTR}) can be approximated as $\mathrm{BTR}(\lambda) $$\,\approx\,$$ N_p(\lambda)/N$, leaving $N_p(\lambda)$ as the leading term. $N_p(\lambda)$ increases with $\lambda$ up to $N$, so when $\lambda$$\,=\,$$ 1$, $N_p(\lambda)$$\,=\,$$ N$. This means that with $\lambda $$\,=\,$$ 1$ we are transmitting more bits than the syntactic case.
\begin{table}[b!]
\vspace{-3mm}
    \renewcommand{\arraystretch}{1.25}
    \centering
    \caption{Optimal threshold values after optimization.}
    \begin{tabular}{c c c c}
    \hline
        $N$ &$\mathbf{\lambda_{opt}}$ &$\mathbf{\alpha(\lambda_{opt})}$ &$\mathbf{BTR(\lambda_{opt})}$\\ 
        \hline
        \textbf{256} & 0.72 & 0.81 & 0.17 \\ 
        \textbf{512} & 0.75 & 0.81 & 0.18 \\ 
        \textbf{1024} & 0.76 & 0.78 & 0.18 \\ 
        \textbf{2048} & 0.78 & 0.76 & 0.2 \\ 
    \hline
    \end{tabular}    
    \label{tab:results}
\end{table}

To balance identification accuracy $\alpha(\lambda)$ and bit transmission ratio $\mathrm{BTR}(\lambda)$, we need to maximize $\alpha(\lambda)$ while minimizing $\mathrm{BTR}(\lambda)$. This can be achieved by maximizing the difference between the two:
\vspace{-.05cm}
\begin{equation}
    \lambda_{opt} = \arg \max_{\lambda} \{ \alpha(\lambda) - \mathrm{BTR}(\lambda) \}.
    \vspace{-.05cm}
\end{equation}
Table~\ref{tab:results} summarizes the results of the optimization process, showing the optimal threshold $\lambda_{opt}$, the corresponding identification accuracy $\alpha(\lambda_{opt})$, and $\mathrm{BTR}(\lambda_{opt})$. Results suggest that smaller feature sets perform better in terms of identification accuracy while larger features sets yields a better BTR. The test case with $5$ features was discarded due to its poor performances. With $5$ features, the transmitter saturates all the possible packets to send before the receiver reaches convergence on a specific class, leading to a very poor performance both in terms of $\mathrm{BTR}$ and identification accuracy.
%
%
%
\section{Conclusions and Further Developments}\label{sec:conclusions}
In this paper, we explored a novel approach to semantic communications by proposing a new communication method based on the identification of semantic features that merges the message identification paradigm with the teacher-apprentice framework, exploiting semantic ambiguity to identify semantic elements with a considerable saving in terms of transmitted bits. Our method enables the receiver to identify the correct semantic element associated with a message with approximately $80\%$ accuracy while using only $18\%$ of the bits required in traditional communication methods and with approximately $90\%$ accuracy using circa $45$$\,-\,$$50\%$ of the bits required in a syntactic communication scheme. This finding demonstrates the potential of semantic communication techniques to reduce bandwidth usage and enhance communication efficiency in the next generation of wireless networks.

Possible future developments for the proposed semantic communication approach include optimizing the feature selection process by determining whether some features are more characteristic than others or exploring advanced feature extraction techniques that leverage different machine learning models depending on the data type. The presented feature identification using inverse distance weighting can become too demanding as the number of features increases. Novel and diverse identification strategies can be explored to tackle this issue. Lastly, a robust method for separating learnable and memorizable data components has yet to be found.
%
%
%
 \section*{Acknowledgment}
The work of Federico F. L. Mariani and Maurizio Magarini was supported in part by the European Union under the Italian National Recovery and Resilience Plan (NRRP) of NextGenerationEU, Partnership on “Telecommunications of the Future” through the Program “RESTART”.
%
\bibliographystyle{IEEEtran}
\bibliography{Bibliography}
\end{document}